\documentclass[floatfix,preprint,aip,jap]{revtex4-2} 
\usepackage{lipsum}
\usepackage{graphicx}
\usepackage{amsmath}
\usepackage{amssymb}
\usepackage{epsfig}
\usepackage{subeqnarray}
\usepackage{color}
\usepackage{bbold}
\usepackage{dsfont}
\usepackage{tikz}
\usepackage{cancel}
\usetikzlibrary{positioning,calc}
\usepackage[version=3]{mhchem}
\usepackage{color}
\usepackage[normalem]{ulem}

\newcommand{\mt}[1]{\mathrm{#1}}

\newcommand{\fig}[1]{Figure~\ref{#1}}
\newcommand{\eq}[1]{Eq.~\ref{#1}}

\newcommand{\E}[1]{$10^{#1}$}

\newcommand{\ie}{\textit{i.e.},~}

\newcommand{\etal}{\textit{et al.~}}

\newcommand{\mum}{~\mt{\mu m}}

\begin{document}

\title{Material Characteristics Governing In-Plane Phonon-Polariton Thermal Conductance}

\author{Jacob Minyard}
\affiliation{School of Mechanical Engineering and Birck Nanotechnology Center, Purdue University, West Lafayette, 47907, IN, USA}
\author{Thomas E. Beechem}
\email{tbeechem@purdue.edu}
\affiliation{School of Mechanical Engineering and Birck Nanotechnology Center, Purdue University, West Lafayette, 47907, IN, USA}
\date{\today}


\begin{abstract}
The material dependence of phonon-polariton based in-plane thermal conductance is investigated by examining systems composed of air and several wurtzite and zinc-blende crystals.  Phonon-polariton based thermal conductance varies by over an order of magnitude ($\sim 0.5-60$  nW/K), which is similar to the variation observed in the materials corresponding bulk thermal conductivity.   Regardless of material, phonon-polaritons exhibit similar thermal conductance to that of phonons when layers become ultrathin ($\sim 10$ nm) suggesting the generality of the effect at these length-scales.  A figure of merit is proposed to explain the large variation of in-plane polariton thermal conductance that is composed entirely of easily predicted and measured optical phonon energies and lifetimes.  Using this figure of merit, in-plane phonon-polariton thermal conductance enlarges with increases in: (1) optical phonon energies, (2) splitting between transverse and longitudinal mode pairs, and (3) phonon lifetimes. 
\end{abstract}
\maketitle

\section{Introduction}

Since the invention of the transistor in 1948, electronic devices have decreased in size and increased in density according to Moore's Law.\cite{moore_2006} Between 2003 and 2022, device scaling increased from \E{8} to \E{11} transistors per chip. \cite{lundstrom_2022}   Size scaling implicit in Moore's Law necessarily enhances heat flux and results in significant device heating. Heating, in turn, accelerates aging, increases parasitic power consumption, detrimentally boosts electromigration, and reduces transistor performance.\cite{landon_2023} At a systems level, the implications of self-heating are profound. Data centers use 1-1.5$\%$ of total worldwide energy production of which nearly a third is dedicated to device cooling.\cite{iea_2022,shehabi_2016,zhang_2021d}

Cooling modern silicon electronics is challenging owing to the size scales involved.  Fin-shaped field effect transistors, finFETs, and gate-all around (GAA) transistors have characteristic features less than 10 nm,\cite{landon_2023} while the metal lines connecting the transistors are $<$20 nm wide.\cite{murdoch_2022}  At these scales, phonons and electrons are significantly less efficient in moving heat.  Phonons in Si, for example, have a predicted thermal conductivity of only $10\%$ relative to bulk at 20 nm, while the resistivity of copper lines increases by at least 2x when wire thickness decreases from 100 to 20 nm.\cite{esfarjani_2011,gall_2020,josell_2009} These reductions are intrinsic to the heat carriers themselves.  They are not the result of extrinsic defects.  Augmenting heat transport at nanoscale therefore necessitates considering alternative approaches and even alternative heat carriers.

Polaritons\textemdash quasiparticles emerging from the hybridization of photons and material dipoles\textemdash are an intriguing possibility for increasing thermal transport in electronic devices. Their heat conductance increases under the same conditions that decrease thermal transport with traditional carriers.  Take, for example, heat transport along a surface caused by the propagation of surface-plasmon\cite{kim_2023b} or surface-phonon\cite{chen_2005} polaritons, which has been termed radiation conduction.\cite{salihoglu_2020} Radiation conduction is much less sensitive to device geometry than either electron or phonon transport.  Under certain circumstances, radiation conduction can even increase with decreasing layer thickness.\cite{wu_2022,chen_2005} This occurs because polaritons move along interfaces, creating long mean free paths for energy transport, while phonons and electrons scatter off interfaces. Radiation conduction increases at higher temperatures, \cite{ordonez-miranda_2014} while traditional heat transport decreases. \cite{ordonez-miranda_2013}  Thanks to these advantages, recent experimental work has shown the significance of radiation conduction as a comparable channel for heat transport relative to phonons and electrons in a variety of materials including:  SiC,\cite{chen_2005,ordonez-miranda_2016} \ce{SiO_2},\cite{tranchant_2019} SiN, \cite{wu_2022} Ti\cite{kim_2023b} and hBN.\cite{baudin_2020}

Despite its potential, the material characteristics that enhance radiation conduction remain relatively unexplored. In response, we examine here the link between optical phonon characteristics and the thermal conductance of radiation conduction stemming from the propagation of surface-phonon polaritons. Simply put, we seek to understand the phonon characteristics that lead to ``big" radiation conduction. This is accomplished by examining the simplest geometry of two semi-infinite planes composed of air on one side and a polar semiconductor with a dielectric function described by its transverse- and longitudinal-optical (TOLO) phonon energies and lifetimes. By surveying radiation conduction for many different materials, clear relationships  between optical phonon properties and polaritonic radiation conduction are deduced. Large optical phonon energies and lifetimes accompanied by sizable splitting between the transverse and longitudinal modes are associated with increases in phonon-polariton driven radiation conduction.

\section{Materials and Methods}
This study quantifies radiation conduction for a number of materials using kinetic theory. All calculations are performed at 300 K unless otherwise noted. Results for gallium arsenide (GaAs), gallium nitride (GaN), and indium antimonide (InSb) are highlighted, since they span a representative range of optical phonon energies and radiation conduction.  The foundation of kinetic theory is the Boltzmann Transport Equation under the single mode relaxation time.\cite{chen_2005} Mathematically, the thermal conductance of the surface phonon-polariton, $\kappa_{SPhP}$, is quantified by integrating over all branches, $s$, and real-parts of the wavevectors, $\beta_r$, of the polariton dispersion as given by:

\begin{equation}
\label{Eq_1}
\kappa_{SPhP} = \frac{1}{4 \pi} \sum_{s} \int_{0}^{\beta_{r,max}} \beta_r \hbar \omega v\Lambda \frac{d f_o}{dT} d\beta_r
\end{equation}
where $\hbar$ is modified Planck's constant, $\omega$ the polariton energy, $v$ its velocity, $\Lambda$ mean free path, and $f_o$ the Bose-Einstein distribution function.  Polaritons decay evanescently away from the interface on which they primarily exist.  Therefore, it is difficult to define a thickness\textemdash and thus an area \textemdash through which the heat moves.  An areal property, thermal conductivity is therefore ill posed when considering radiation conduction in the same way it is when considering transport through 2D-solids.\cite{wu_2017a}  Consequently, \eq{Eq_1} has been derived using a two-dimensional density of states.  As such, $k_{SPhP}$ quantifies conductance rather than a conductivity and has units of $\left[\frac{W}{K}\right]$ instead of the typical conductivity $\left[\frac{W}{m \cdot K}\right]$. The conductance approach is consistent with established means of comparing the thermal properties of ultrathin solids.\cite{wu_2017a} 

The polariton dispersion is determined by the boundary conditions of Maxwell's equations and is therefore dependent upon the optical properties of the materials on either side of the interface.  For the polar material, the optical properties are described by a dielectric function defined by the energies and lifetimes of the transverse- (TO) and longitudinal optical (LO) phonons via:
\begin{equation}
\label{Eq_2}
\epsilon(\omega) = \epsilon_\infty \left(1 + \sum_{i}^{n} \frac{\omega_{LO,i}^2 - \omega_{TO,i}^2}{\omega_{TO,i}^2 - \omega^2 - i \omega \Gamma_i}\right)
\end{equation}
where $\epsilon_\infty$ is the high-frequency permittivity of the solid, $\omega_{TO(LO)}$ is the frequency of the transverse-optical (longitudinal-optical) phonons and $\Gamma$ is the mean phonon lifetime. Phonon energies and lifetimes were taken from the literature.\cite{togo_2015,jarzembski_2020}  To compare intrinsic material responses, phonon lifetimes from first-principles calculations were employed.\cite{togo_2015} Values for these parameters are tabulated in the Supplemental Material.

Our model consists of two semi-infinite planes made up of air on the upper plane and the polar dielectric having a dielectric permittivity of the form of \eq{Eq_2} as the lower plane. This arrangement is shown schematically in the upper panel of \fig{Fig_1}. For this simple arrangement, the polariton dispersion can be analytically determined and is given by:
\begin{equation}
\label{Eq_3}
\beta = k_0 \sqrt{\frac{\epsilon_1 \epsilon_2}{\epsilon_1 + \epsilon_2}}
\end{equation}
where $k_0=\omega/c$ is the wavevector of the incident light in vacuum and $\epsilon_{1,2}$ are the permittivities of the materials in the model.  A slightly more involved relation can also be found for birefringent materials and is provided in the Supplemental Material. 

As the dielectric function is a complex quantity, the resulting wavevectors are likewise complex.  The real-part of the wavevector ($\beta_r$) provides the in-plane momentum for the species and is therefore utilized in \eq{Eq_1} defining energy transport. Only modes where $\beta_r > k_0$ are considered in the analysis corresponding to the frequency range of $\omega_{TO}\leq \omega \leq \omega_{LO}$. Beyond this range, Brewster modes exist above the light-line but are not localized to the interface (\ie not true surface modes). They have a finite wavevector pointing orthogonal to the temperature gradient and are therefore assumed to be of secondary importance to radiation conduction.\cite{novotny_2012,borstel_1977,chen_2007a}

The imaginary portion of the wavevector ($\beta_i$) is related to the propagation length of the polariton through:
\begin{equation}
\label{Eq_4}
\Lambda = \frac{1}{\beta_i}
\end{equation}
It is therefore a reasonable surrogate for the mean free path and is used as such. As the dispersion of \eq{Eq_3} is derived assuming an infinite lateral plane making up the interface, $\Lambda$ does not take into account any finite size effects that may impact the scattering of the phonon-polariton or any other extrinsic effect.  \eq{Eq_4} differs by a factor of two relative to that commonly employed in similar treatments of radiation conduction;\cite{chen_2005a} it is used here owing to its correlation with experimental results.\cite{huber_2005} 

\section{Results and Discussion}

\begin{figure}[htbp]
\centering
\includegraphics[scale = 0.85]{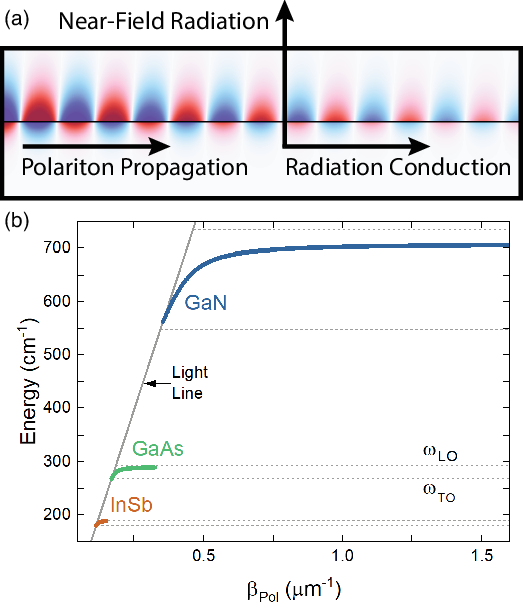}
\caption{(a) Model schematic. Near-field radiation occurs perpendicular to polariton propagation whereas radiation conduction occurs parallel to polariton propagation. The system is composed of two infinite planes consisting of air (top) and a polar semiconductor (bottom). The phonon-polariton is localized to the interface as shown here by the simulated electric-field contours for the mode. (b) Phonon-polariton dispersion for GaN, GaAs, and InSb. The horizontal lines show their respective LO and TO phonons; LO phonons reside at higher energies.}
\label{Fig_1}
\end{figure}

\fig{Fig_1} presents the resulting phonon-polariton dispersion for GaN, GaAs, and InSb. These materials are highlighted from among the more than twenty analyzed because their vibrational characteristics span much of the range observed for polar crystalline solids.  For each material, the phonon-polariton dispersion branches off of the so-called light line (diagonal line in \fig{Fig_1}).  Having a slope that is comparable to the light-line, phonon-polaritons are characterized by extremely high velocity (on the order of $10^7$ $\frac{m}{s}$), but much smaller momentum, than the phonons from which they derive. The dispersion is also bounded by the energies of the LO and TO phonons. Therefore, materials with large optical phonon splitting have more phase space to create polaritons and thus a greater total population.

Radiation conduction from phonon-polaritons necessarily depends upon the characteristics of the LO and TO phonons via \eq{Eq_2}. Traditional thermal conductivity is dependent upon all phonons existing within the solid.  Recognizing this overlap, there is significant correlation between polariton conductance and bulk thermal conductivity as is apparent upon inspection of Figure 2.  Increasing polaritonic conductance is correlated with higher thermal conductivity.   However, the correlation between polaritonic conductance and thermal conductivity is not complete, since phonon thermal conductivity is determined by the entirety of the Brillouin zone, whereas polaritonic conductance is driven by only $\Gamma$-point optical phonons. Examining the characteristics of $\Gamma$-point phonons, therefore, is a means of understanding the characteristics of phonon-driven radiation conduction. 

\begin{figure}[htbp]
\centering
\includegraphics[width=88 mm]{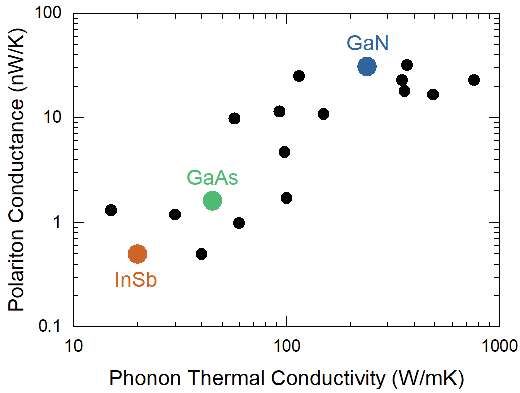}
\caption{Correlation between phonon thermal conductivity and phonon-polariton driven radiation conduction at 300 K for polar semiconductors with dielectric functions described under the TOLO-formalism of \eq{Eq_4}. Phonon thermal conductivity values are taken from Togo \etal \cite{togo_2015}. Materials are individually labeled in a complementary version of this figure provided within the Supplemental Material.} 
\label{Fig_2}
\end{figure}

\begin{figure}[htbp]
\centering
\includegraphics[width=160 mm]{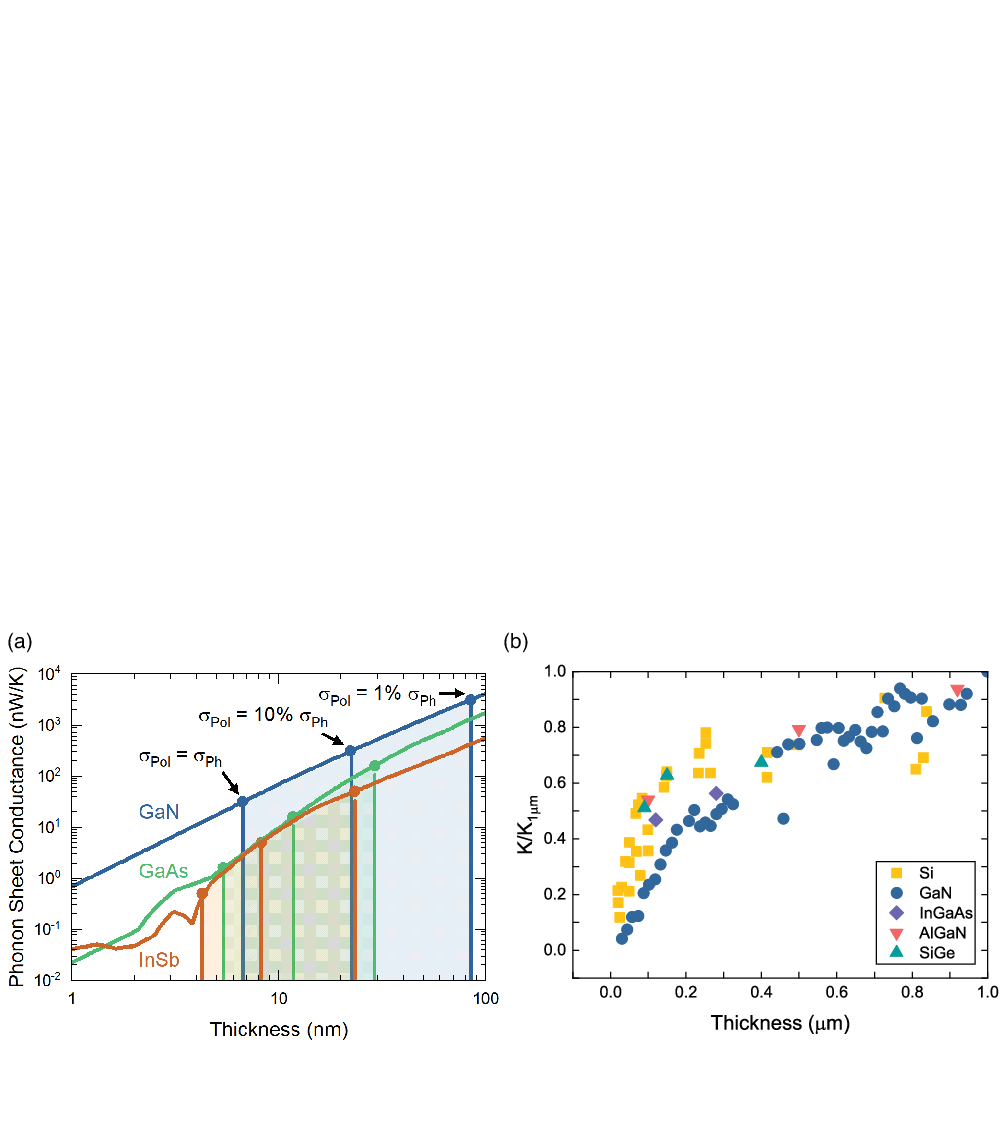}
\caption{(a) Phonon conductance versus thickness for \ce{InSb}, \ce{GaAs}, and \ce{GaN}. Shaded regions bound those thicknesses in which the radiation conduction contribution from phonon polaritons spans 1 to 100$\%$ that of the phonon value. (b) Measured thickness dependent thermal conductivity at room temperature normalized relative to its value at a thickness of $1 \mum$. Experimental data is drawn from Si\cite{marconnet_2013}, GaN\cite{ziade_2017}, InGaAs\cite{seyf_2017}, AlGaN\cite{song_2021a}, SiGe.\cite{cheaito_2012}}
\label{Fig_3}
\end{figure}

The remainder of this paper quantifies the relative amount of heat moved by radiation conduction and then seeks to understand the underlying phonon characteristics by which radiation conduction can be maximized, using three case studies (InSb, GaAs, and GaN) with  small, medium, and large phonon and polariton conductances respectively (see \fig{Fig_2}).  To quantify the relation between phonon and polariton conductances, the size-dependent phonon conductance was calculated using previously reported mean free path spectra and thermal conductivity accumulation functions for InSb, GaAs, and GaN.\cite{lee_2014c,luo_2013a,beechem_2016a} The size-dependent phonon thermal conductance $\sigma_{ph}$ was calculated via:\cite{wu_2017a} 
\begin{equation}\label{}
\sigma_{ph} = \kappa(t)t
\end{equation}
where $\kappa(t)$ is the phonon thermal conductivity at a given thickness $t$, deduced by multiplying the bulk thermal conductivity $\kappa_{bulk}$ by its accumulation function. The results are depicted in \fig{Fig_3} where phonon conductance is compared to polariton conductance as a function of slab thickness for in-plane transport.  It should be noted that the polaritonic dispersion\textemdash and thus the polaritonic conductance\textemdash will be changed as the film approaches sub-100 nm length scales owing to interaction of the fields between the surfaces.\cite{chen_2005a}  The effect is ignored here, however, to allow for material comparisons apart from geometric effects.  

As \fig{Fig_3}(a) shows, phonon conductance and polariton-based conductance are comparable at thicknesses of 10 nm regardless of the large differences in bulk thermal conductivity between all materials. This occurs because phonon thermal conductivity reduces with thickness at roughly comparable rates regardless of its bulk value. This is indicated in \fig{Fig_3}(b), which plots previously-reported thickness-dependent thermal conductivities measured at room temperature for several materials that are normalized relative to their value at a thickness of $1 \mum$. Although the examined materials exhibit bulk thermal conductivities that vary by a factor exceeding 100, the rate of reduction in thermal conductivity is similar for the entire set of materials. Along with the correlation observed in \fig{Fig_2}, this trend explains why phonon-polariton based radiation conduction becomes comparable to that of traditional phonon conduction at similar length scales for very different materials. It should therefore be presumed that radiation conduction mediated by phonon-polaritons is intrinsically capable of playing a significant role in heat transport for polar materials with lengths approaching 10 nm.

\begin{figure}[htbp]
\centering
\includegraphics[width=85 mm]{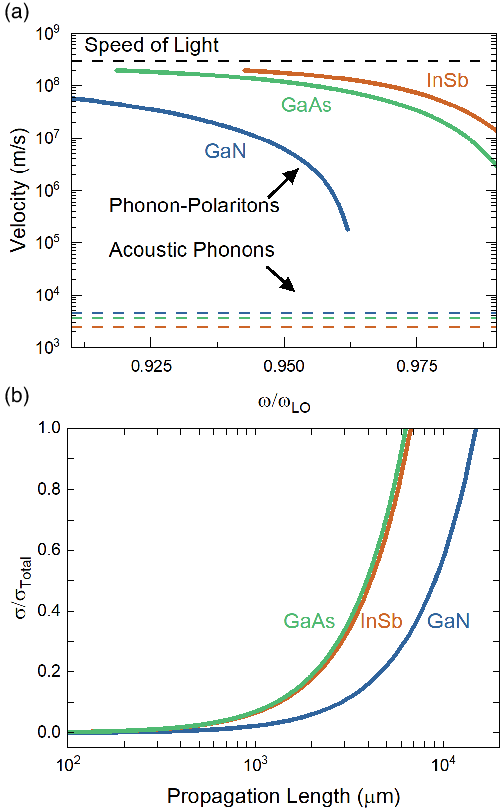}
\caption{(a) Velocities of phonon-polaritons and acoustic phonons for InSb, GaAs, and GaN. Phonon-velocities acquired from data reported in InSb\cite{slutsky_1959}, GaAs\cite{blakemore_1982}, and GaN\cite{wright_1997}  (b) Accumulation of phonon-polariton radiation conductance versus polariton propagation distance calculated from Eq. 5.  The majority of heat is transported by polaritons having a propagation length greater than a 1 mm.}
\label{Fig_4}
\end{figure}

 Phonon-polaritons move appreciable amounts of energy for three reasons. First, phonon-polaritons couple to optical phonons and thus have larger energies than the acoustic phonons that dominate standard phonon conduction. Second, their photonic character permits group velocities closer to the speed of light, around ten times the speed of phonons, which move at the speed of sound. This is indicated by the dispersion curves in Figure 1, wherein the polaritonic dispersion lies close to the light line, implying a velocity close to that of light as can be seen explicitly in \fig{Fig_4}(a), where the phonon-polariton velocity for InSb, GaAs, and GaN is plotted.  
The group velocities of the phonon-polaritons in InSb, GaAs, and GaN exceed $10^7~\mt{m/s}$ throughout their dispersion. This is four orders of magnitude greater than acoustic phonon speeds, which typically are on the order of $10^3~\mt{m/s}$. Finally, phonon-polaritons move exceptionally long distances before scattering. Figure 4b shows that the vast majority of radiation conduction is mediated by phonon-polaritons with propagation lengths greater than 1 mm, consistent with previous reports and far longer than the mean free path of phonons. \cite{chen_2005a,ordonez-miranda_2013}  

\begin{figure}[htbp]
\centering
\includegraphics[width=85 mm]{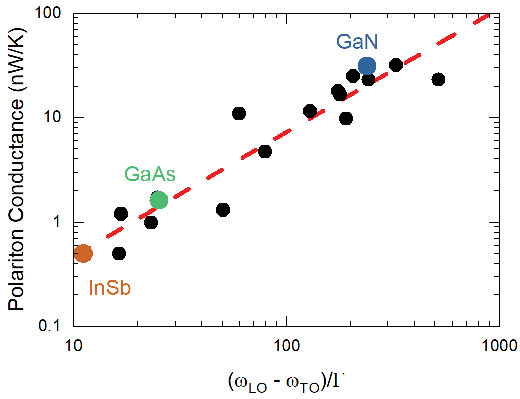}
\caption{Phonon-polariton radiation conductance versus Figure of Merit (FoM) (see Eq. 6) for each of the examined materials.  Over 80$\%$ of the plot's variance is described through the FoM, which derives from the characteristic features of energy, speed, and propagation of the phonon-polariton. Materials are individually labeled in a complementary version of this figure provided within the Supplemental Material.}
\label{Fig_5}
\end{figure}

The dielectric function of \eq{Eq_2} ultimately defines the dispersion and propagation length of the phonon-polaritons and is dependent upon only phonon energies and lifetimes. This suggests that phonon-characteristics can be used to define a figure of merit ($FoM$) to compare materials in their ability for radiation conduction. We define the $FoM$ for phonon-polariton driven radiation conduction in \eq{eq_FOM}: 
\begin{equation}\label{eq_FOM}
FoM = \frac{\omega_{LO}-\omega_{TO}}{\Gamma}
\end{equation}
\fig{Fig_5} plots the predicted conductance for all materials versus this value.

The proposed figure of merit strongly links the magnitude of polariton conductance with $\Gamma$-point phonon energies and lifetimes.  The strong correlation can be understood by examining the $FoM$ in light of the characteristics of radiation conduction.  The numerator of \eq{eq_FOM} ($\omega_{LO}-\omega_{TO}$) describes the energy of the phonon-polaritons and the phase-space available for their creation based on the difference between the corresponding TO- and LO-modes. A larger difference implies ``more" phonon-polaritons of higher energy. The denominator ($\Gamma$) is a quantification of loss induced by the phonons and thus is a measure of polariton propagation. A higher value of $\Gamma$ corresponds to smaller propagation and thus less efficient transport. The figure of merit does not take into account the scaling of polariton population that increases as the thermal energy (208 $\mathrm{cm^{-1}}$ at 300 K) approaches the TO and LO-phonon energies.  Even with this omission, the simple ratio explains over 80$\%$ of the variance between materials.

\begin{figure}[htbp]
\centering
\includegraphics[width=85 mm]{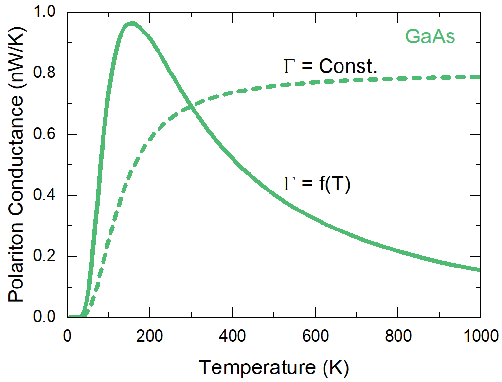}
\caption{Temperature dependence of phonon-polariton radiation conduction in GaAs when considering both (solid line) temperature dependent and (dash) constant phonon linewidth. Radiation conduction varies with temperature as does the phonon linewidth.}
\label{Fig_6}
\end{figure}

As the phonon lifetime plays a central role in determining the magnitude of radiation conduction. The temperature dependence of phonon lifetime will, therefore, dictate the temperature dependence of radiation conduction. Due to anharmonicity, the phonon linewidth increases with temperature in a manner that impacts radiation conduction. \cite{beechem_2008a,yang_2020a,liarokapis_1984,verma_1995} To account for this fact while removing any extrinsic causes, the temperature-dependent polariton conductance for GaAs was calculated using the first principles estimations for phonon lifetime reported in Yang \etal \cite{yang_2020a} Temperature-dependent changes in phonon energies were not accounted for since they are comparatively small relative to the variation in linewidth. \fig{Fig_6} plots the resulting temperature-dependent phonon-polariton radiation conduction for GaAs.  

When considering a temperature dependent phonon lifetime, a maximum is observed in the polariton conductance near 150 K. This maximum is distinct from the Umklapp hump of GaAs that occurs near 20 K of phonon conduction\cite{carlson_1965} since the radiation conduction is dictated by the balance in phonon-polariton population and scattering rather than phonon population and scattering. The decrease in radiation conduction with temperature above room temperature is contrary to previous reports. \cite{salihoglu_2020,wu_2020} There are two possibilities to account for the discrepancy. Extrinsic polariton scattering from defects or size effects could dominate the intrinsic temperature-dependent changes in phonon lifetime. It is also possible that the discrepancy indicates a limitation in the ability of kinetic theory to fully describe radiation conduction, suggesting that approaches emphasizing fluctuational electrodynamics may be more relevant.\cite{chen_2010b,salihoglu_2020,kim_2023b} Regardless of causation, there remains much to explore regarding the effects of temperature and other extrinsic factors on radiation conduction.  

\section{Conclusions}
In-plane phonon-polariton thermal conductance\textemdash termed radiation conduction\textemdash takes on a value that is comparable to phonon conduction for ultrathin solids ($\sim 10 $ nm) regardless of the material's bulk thermal conductivity. This is because phonon-polaritons have comparatively high energy combined with extreme velocities and long intrinsic mean-free paths.  These characteristics derive from the dispersion of the phonon-polariton that itself is primarily determined by the $\Gamma-$point optical phonon energies and lifetimes. A figure of merit ($FoM = \tfrac{(\omega_{LO} - \omega_{TO})}{\Gamma}$) was established leveraging these optical phonon characteristics that was highly correlative with the over 10x variation in radiation conduction observed for the 20 different wurtzite and zinc-blende materials examined here. This figure of merit highlights that radiation conduction increases for high-energy phonons having long lifetimes and can be used identify promising materials to cultivate radiation conduction.

\section{Supplementary Material}
Optical properties and thermal conductivity values utilized within the analysis presented here is provided in the Supplementary Material along with the expression describing the surface-phonon polariton dispersion for air atop a birefringent material. \fig{Fig_2} and 5 are reproduced with each material labelled individually.

\section{Data Availability}
Code used to produce the figures in this manuscript and underlying data is available upon reasonable request to the corresponding author. 

%



\begin{thebibliography}{45}%
\makeatletter
\providecommand \@ifxundefined [1]{%
 \@ifx{#1\undefined}
}%
\providecommand \@ifnum [1]{%
 \ifnum #1\expandafter \@firstoftwo
 \else \expandafter \@secondoftwo
 \fi
}%
\providecommand \@ifx [1]{%
 \ifx #1\expandafter \@firstoftwo
 \else \expandafter \@secondoftwo
 \fi
}%
\providecommand \natexlab [1]{#1}%
\providecommand \enquote  [1]{``#1''}%
\providecommand \bibnamefont  [1]{#1}%
\providecommand \bibfnamefont [1]{#1}%
\providecommand \citenamefont [1]{#1}%
\providecommand \href@noop [0]{\@secondoftwo}%
\providecommand \href [0]{\begingroup \@sanitize@url \@href}%
\providecommand \@href[1]{\@@startlink{#1}\@@href}%
\providecommand \@@href[1]{\endgroup#1\@@endlink}%
\providecommand \@sanitize@url [0]{\catcode `\\12\catcode `\$12\catcode
  `\&12\catcode `\#12\catcode `\^12\catcode `\_12\catcode `\%12\relax}%
\providecommand \@@startlink[1]{}%
\providecommand \@@endlink[0]{}%
\providecommand \url  [0]{\begingroup\@sanitize@url \@url }%
\providecommand \@url [1]{\endgroup\@href {#1}{\urlprefix }}%
\providecommand \urlprefix  [0]{URL }%
\providecommand \Eprint [0]{\href }%
\providecommand \doibase [0]{https://doi.org/}%
\providecommand \selectlanguage [0]{\@gobble}%
\providecommand \bibinfo  [0]{\@secondoftwo}%
\providecommand \bibfield  [0]{\@secondoftwo}%
\providecommand \translation [1]{[#1]}%
\providecommand \BibitemOpen [0]{}%
\providecommand \bibitemStop [0]{}%
\providecommand \bibitemNoStop [0]{.\EOS\space}%
\providecommand \EOS [0]{\spacefactor3000\relax}%
\providecommand \BibitemShut  [1]{\csname bibitem#1\endcsname}%
\let\auto@bib@innerbib\@empty
\bibitem [{\citenamefont {Moore}(2006)}]{moore_2006}%
  \BibitemOpen
  \bibfield  {author} {\bibinfo {author} {\bibfnamefont {G.~E.}\ \bibnamefont
  {Moore}},\ }\bibfield  {title} {\enquote {\bibinfo {title} {Cramming more
  components onto integrated circuits, {{Reprinted}} from {{Electronics}},
  volume 38, number 8, {{April}} 19, 1965, pp.114 ff.}}\ }\href
  {https://doi.org/10.1109/N-SSC.2006.4785860} {\bibfield  {journal} {\bibinfo
  {journal} {IEEE Solid-State Circuits Soc. Newsl.}\ }\textbf {\bibinfo
  {volume} {11}},\ \bibinfo {pages} {33--35} (\bibinfo {year}
  {2006})}\BibitemShut {NoStop}%
\bibitem [{\citenamefont {Lundstrom}\ and\ \citenamefont
  {Alam}(2022)}]{lundstrom_2022}%
  \BibitemOpen
  \bibfield  {author} {\bibinfo {author} {\bibfnamefont {M.~S.}\ \bibnamefont
  {Lundstrom}}\ and\ \bibinfo {author} {\bibfnamefont {M.~A.}\ \bibnamefont
  {Alam}},\ }\bibfield  {title} {\enquote {\bibinfo {title} {Moore's law:
  {{The}} journey ahead},}\ }\href {https://doi.org/10.1126/science.ade2191}
  {\bibfield  {journal} {\bibinfo  {journal} {Science}\ }\textbf {\bibinfo
  {volume} {378}},\ \bibinfo {pages} {722--723} (\bibinfo {year}
  {2022})}\BibitemShut {NoStop}%
\bibitem [{\citenamefont {Landon}\ \emph {et~al.}(2023)\citenamefont {Landon},
  \citenamefont {Jiang}, \citenamefont {Pantuso}, \citenamefont {Meric},
  \citenamefont {Komeyli}, \citenamefont {Hicks},\ and\ \citenamefont
  {Schroeder}}]{landon_2023}%
  \BibitemOpen
  \bibfield  {author} {\bibinfo {author} {\bibfnamefont {C.}~\bibnamefont
  {Landon}}, \bibinfo {author} {\bibfnamefont {L.}~\bibnamefont {Jiang}},
  \bibinfo {author} {\bibfnamefont {D.}~\bibnamefont {Pantuso}}, \bibinfo
  {author} {\bibfnamefont {I.}~\bibnamefont {Meric}}, \bibinfo {author}
  {\bibfnamefont {K.}~\bibnamefont {Komeyli}}, \bibinfo {author} {\bibfnamefont
  {J.}~\bibnamefont {Hicks}},\ and\ \bibinfo {author} {\bibfnamefont
  {D.}~\bibnamefont {Schroeder}},\ }\bibfield  {title} {\enquote {\bibinfo
  {title} {Localized thermal effects in {{Gate-all-around}} devices},}\ }in\
  \href {https://doi.org/10.1109/IRPS48203.2023.10117903} {\emph {\bibinfo
  {booktitle} {2023 {{IEEE Int}}. {{Reliab}}. {{Phys}}. {{Symp}}. {{IRPS}}}}}\
  (\bibinfo {year} {2023})\ pp.\ \bibinfo {pages} {1--5}\BibitemShut {NoStop}%
\bibitem [{\citenamefont {IEA}(2022)}]{iea_2022}%
  \BibitemOpen
  \bibfield  {author} {\bibinfo {author} {\bibnamefont {IEA}},\ }\href@noop {}
  {\enquote {\bibinfo {title} {Data {{Centres}} and {{Data Transmission
  Networks}} \textendash{} {{Analysis}}},}\ }\bibinfo {type} {Tech. Rep.}\
  (\bibinfo {year} {2022})\BibitemShut {NoStop}%
\bibitem [{\citenamefont {Shehabi}\ \emph {et~al.}(2016)\citenamefont
  {Shehabi}, \citenamefont {Smith}, \citenamefont {Sartor}, \citenamefont
  {Brown}, \citenamefont {Herrlin}, \citenamefont {Koomey}, \citenamefont
  {Masanet}, \citenamefont {Horner}, \citenamefont {Azevedo},\ and\
  \citenamefont {Lintner}}]{shehabi_2016}%
  \BibitemOpen
  \bibfield  {author} {\bibinfo {author} {\bibfnamefont {A.}~\bibnamefont
  {Shehabi}}, \bibinfo {author} {\bibfnamefont {S.}~\bibnamefont {Smith}},
  \bibinfo {author} {\bibfnamefont {D.}~\bibnamefont {Sartor}}, \bibinfo
  {author} {\bibfnamefont {R.}~\bibnamefont {Brown}}, \bibinfo {author}
  {\bibfnamefont {M.}~\bibnamefont {Herrlin}}, \bibinfo {author} {\bibfnamefont
  {J.}~\bibnamefont {Koomey}}, \bibinfo {author} {\bibfnamefont
  {E.}~\bibnamefont {Masanet}}, \bibinfo {author} {\bibfnamefont
  {N.}~\bibnamefont {Horner}}, \bibinfo {author} {\bibfnamefont
  {I.}~\bibnamefont {Azevedo}},\ and\ \bibinfo {author} {\bibfnamefont
  {W.}~\bibnamefont {Lintner}},\ }\href {https://doi.org/10.2172/1372902}
  {\enquote {\bibinfo {title} {United {{States Data Center Energy Usage
  Report}}},}\ }\bibinfo {type} {Tech. Rep.}\ \bibinfo {number} {LBNL--1005775,
  1372902}\ (\bibinfo {year} {2016})\BibitemShut {NoStop}%
\bibitem [{\citenamefont {Zhang}\ \emph {et~al.}(2021)\citenamefont {Zhang},
  \citenamefont {Meng}, \citenamefont {Hong}, \citenamefont {Zhan},
  \citenamefont {Liu}, \citenamefont {Dong}, \citenamefont {Bai}, \citenamefont
  {Niu},\ and\ \citenamefont {Deen}}]{zhang_2021d}%
  \BibitemOpen
  \bibfield  {author} {\bibinfo {author} {\bibfnamefont {Q.}~\bibnamefont
  {Zhang}}, \bibinfo {author} {\bibfnamefont {Z.}~\bibnamefont {Meng}},
  \bibinfo {author} {\bibfnamefont {X.}~\bibnamefont {Hong}}, \bibinfo {author}
  {\bibfnamefont {Y.}~\bibnamefont {Zhan}}, \bibinfo {author} {\bibfnamefont
  {J.}~\bibnamefont {Liu}}, \bibinfo {author} {\bibfnamefont {J.}~\bibnamefont
  {Dong}}, \bibinfo {author} {\bibfnamefont {T.}~\bibnamefont {Bai}}, \bibinfo
  {author} {\bibfnamefont {J.}~\bibnamefont {Niu}},\ and\ \bibinfo {author}
  {\bibfnamefont {M.~J.}\ \bibnamefont {Deen}},\ }\bibfield  {title} {\enquote
  {\bibinfo {title} {A survey on data center cooling systems: {{Technology}},
  power consumption modeling and control strategy optimization},}\ }\href
  {https://doi.org/10.1016/j.sysarc.2021.102253} {\bibfield  {journal}
  {\bibinfo  {journal} {Journal of Systems Architecture}\ }\textbf {\bibinfo
  {volume} {119}},\ \bibinfo {pages} {102253} (\bibinfo {year}
  {2021})}\BibitemShut {NoStop}%
\bibitem [{\citenamefont {Murdoch}\ \emph {et~al.}(2022)\citenamefont
  {Murdoch}, \citenamefont {O'Toole}, \citenamefont {Marti}, \citenamefont
  {Pokhrel}, \citenamefont {Tsvetanova}, \citenamefont {Decoster},
  \citenamefont {Kundu}, \citenamefont {Oniki}, \citenamefont {Thiam},
  \citenamefont {Le}, \citenamefont {Varela~Pedreira}, \citenamefont
  {Lesniewska}, \citenamefont {{Martinez-Alanis}}, \citenamefont {Park},\ and\
  \citenamefont {Tokei}}]{murdoch_2022}%
  \BibitemOpen
  \bibfield  {author} {\bibinfo {author} {\bibfnamefont {G.}~\bibnamefont
  {Murdoch}}, \bibinfo {author} {\bibfnamefont {M.}~\bibnamefont {O'Toole}},
  \bibinfo {author} {\bibfnamefont {G.}~\bibnamefont {Marti}}, \bibinfo
  {author} {\bibfnamefont {A.}~\bibnamefont {Pokhrel}}, \bibinfo {author}
  {\bibfnamefont {D.}~\bibnamefont {Tsvetanova}}, \bibinfo {author}
  {\bibfnamefont {S.}~\bibnamefont {Decoster}}, \bibinfo {author}
  {\bibfnamefont {S.}~\bibnamefont {Kundu}}, \bibinfo {author} {\bibfnamefont
  {Y.}~\bibnamefont {Oniki}}, \bibinfo {author} {\bibfnamefont
  {A.}~\bibnamefont {Thiam}}, \bibinfo {author} {\bibfnamefont
  {Q.}~\bibnamefont {Le}}, \bibinfo {author} {\bibfnamefont {O.}~\bibnamefont
  {Varela~Pedreira}}, \bibinfo {author} {\bibfnamefont {A.}~\bibnamefont
  {Lesniewska}}, \bibinfo {author} {\bibfnamefont {G.}~\bibnamefont
  {{Martinez-Alanis}}}, \bibinfo {author} {\bibfnamefont {S.}~\bibnamefont
  {Park}},\ and\ \bibinfo {author} {\bibfnamefont {{\relax Zs}.}~\bibnamefont
  {Tokei}},\ }\bibfield  {title} {\enquote {\bibinfo {title} {First
  demonstration of {{Two Metal Level Semi-damascene Interconnects}} with
  {{Fully Self-aligned Vias}} at {{18MP}}},}\ }in\ \href
  {https://doi.org/10.1109/VLSITechnologyandCir46769.2022.9830150} {\emph
  {\bibinfo {booktitle} {2022 {{IEEE Symp}}. {{VLSI Technol}}. {{Circuits VLSI
  Technol}}. {{Circuits}}}}}\ (\bibinfo {year} {2022})\ pp.\ \bibinfo {pages}
  {1--2}\BibitemShut {NoStop}%
\bibitem [{\citenamefont {Esfarjani}, \citenamefont {Chen},\ and\ \citenamefont
  {Stokes}(2011)}]{esfarjani_2011}%
  \BibitemOpen
  \bibfield  {author} {\bibinfo {author} {\bibfnamefont {K.}~\bibnamefont
  {Esfarjani}}, \bibinfo {author} {\bibfnamefont {G.}~\bibnamefont {Chen}},\
  and\ \bibinfo {author} {\bibfnamefont {H.~T.}\ \bibnamefont {Stokes}},\
  }\bibfield  {title} {\enquote {\bibinfo {title} {Heat transport in silicon
  from first-principles calculations},}\ }\href@noop {} {\bibfield  {journal}
  {\bibinfo  {journal} {Phys. Rev. B}\ }\textbf {\bibinfo {volume} {84}},\
  \bibinfo {pages} {085204} (\bibinfo {year} {2011})}\BibitemShut {NoStop}%
\bibitem [{\citenamefont {Gall}(2020)}]{gall_2020}%
  \BibitemOpen
  \bibfield  {author} {\bibinfo {author} {\bibfnamefont {D.}~\bibnamefont
  {Gall}},\ }\bibfield  {title} {\enquote {\bibinfo {title} {The search for the
  most conductive metal for narrow interconnect lines},}\ }\href
  {https://doi.org/10.1063/1.5133671} {\bibfield  {journal} {\bibinfo
  {journal} {Journal of Applied Physics}\ }\textbf {\bibinfo {volume} {127}},\
  \bibinfo {pages} {050901} (\bibinfo {year} {2020})}\BibitemShut {NoStop}%
\bibitem [{\citenamefont {Josell}, \citenamefont {Brongersma},\ and\
  \citenamefont {T{\H o}kei}(2009)}]{josell_2009}%
  \BibitemOpen
  \bibfield  {author} {\bibinfo {author} {\bibfnamefont {D.}~\bibnamefont
  {Josell}}, \bibinfo {author} {\bibfnamefont {S.~H.}\ \bibnamefont
  {Brongersma}},\ and\ \bibinfo {author} {\bibfnamefont {Z.}~\bibnamefont {T{\H
  o}kei}},\ }\bibfield  {title} {\enquote {\bibinfo {title} {Size-{{Dependent
  Resistivity}} in {{Nanoscale Interconnects}}},}\ }\href
  {https://doi.org/10.1146/annurev-matsci-082908-145415} {\bibfield  {journal}
  {\bibinfo  {journal} {Annu. Rev. Mater. Res.}\ }\textbf {\bibinfo {volume}
  {39}},\ \bibinfo {pages} {231--254} (\bibinfo {year} {2009})}\BibitemShut
  {NoStop}%
\bibitem [{\citenamefont {Kim}\ \emph {et~al.}(2023)\citenamefont {Kim},
  \citenamefont {Choi}, \citenamefont {Cho}, \citenamefont {Lim},\ and\
  \citenamefont {Lee}}]{kim_2023b}%
  \BibitemOpen
  \bibfield  {author} {\bibinfo {author} {\bibfnamefont {D.-m.}\ \bibnamefont
  {Kim}}, \bibinfo {author} {\bibfnamefont {S.}~\bibnamefont {Choi}}, \bibinfo
  {author} {\bibfnamefont {J.}~\bibnamefont {Cho}}, \bibinfo {author}
  {\bibfnamefont {M.}~\bibnamefont {Lim}},\ and\ \bibinfo {author}
  {\bibfnamefont {B.~J.}\ \bibnamefont {Lee}},\ }\bibfield  {title} {\enquote
  {\bibinfo {title} {Boosting {{Thermal Conductivity}} by {{Surface Plasmon
  Polaritons Propagating}} along a {{Thin Ti Film}}},}\ }\href
  {https://doi.org/10.1103/PhysRevLett.130.176302} {\bibfield  {journal}
  {\bibinfo  {journal} {Phys. Rev. Lett.}\ }\textbf {\bibinfo {volume} {130}},\
  \bibinfo {pages} {176302} (\bibinfo {year} {2023})}\BibitemShut {NoStop}%
\bibitem [{\citenamefont {Chen}(2005)}]{chen_2005}%
  \BibitemOpen
  \bibfield  {author} {\bibinfo {author} {\bibfnamefont {G.}~\bibnamefont
  {Chen}},\ }\href@noop {} {\emph {\bibinfo {title} {Nanoscale {{Energy
  Transport}} and {{Conversion}}}}}\ (\bibinfo  {publisher} {{Oxford University
  Press}},\ \bibinfo {address} {{New York, NY}},\ \bibinfo {year}
  {2005})\BibitemShut {NoStop}%
\bibitem [{\citenamefont {Salihoglu}\ \emph {et~al.}(2020)\citenamefont
  {Salihoglu}, \citenamefont {Iyer}, \citenamefont {Taniguchi}, \citenamefont
  {Watanabe}, \citenamefont {Ye},\ and\ \citenamefont {Xu}}]{salihoglu_2020}%
  \BibitemOpen
  \bibfield  {author} {\bibinfo {author} {\bibfnamefont {H.}~\bibnamefont
  {Salihoglu}}, \bibinfo {author} {\bibfnamefont {V.}~\bibnamefont {Iyer}},
  \bibinfo {author} {\bibfnamefont {T.}~\bibnamefont {Taniguchi}}, \bibinfo
  {author} {\bibfnamefont {K.}~\bibnamefont {Watanabe}}, \bibinfo {author}
  {\bibfnamefont {P.~D.}\ \bibnamefont {Ye}},\ and\ \bibinfo {author}
  {\bibfnamefont {X.}~\bibnamefont {Xu}},\ }\bibfield  {title} {\enquote
  {\bibinfo {title} {Energy {{Transport}} by {{Radiation}} in {{Hyperbolic
  Material Comparable}} to {{Conduction}}},}\ }\href
  {https://doi.org/10.1002/adfm.201905830} {\bibfield  {journal} {\bibinfo
  {journal} {Adv. Funct. Mater.}\ }\textbf {\bibinfo {volume} {30}},\ \bibinfo
  {pages} {1905830} (\bibinfo {year} {2020})}\BibitemShut {NoStop}%
\bibitem [{\citenamefont {Wu}\ \emph {et~al.}(2022)\citenamefont {Wu},
  \citenamefont {Duan}, \citenamefont {Ma}, \citenamefont {Ou}, \citenamefont
  {Li}, \citenamefont {{Alonso-Gonz{\'a}lez}}, \citenamefont {Caldwell},\ and\
  \citenamefont {Bao}}]{wu_2022}%
  \BibitemOpen
  \bibfield  {author} {\bibinfo {author} {\bibfnamefont {Y.}~\bibnamefont
  {Wu}}, \bibinfo {author} {\bibfnamefont {J.}~\bibnamefont {Duan}}, \bibinfo
  {author} {\bibfnamefont {W.}~\bibnamefont {Ma}}, \bibinfo {author}
  {\bibfnamefont {Q.}~\bibnamefont {Ou}}, \bibinfo {author} {\bibfnamefont
  {P.}~\bibnamefont {Li}}, \bibinfo {author} {\bibfnamefont {P.}~\bibnamefont
  {{Alonso-Gonz{\'a}lez}}}, \bibinfo {author} {\bibfnamefont {J.~D.}\
  \bibnamefont {Caldwell}},\ and\ \bibinfo {author} {\bibfnamefont
  {Q.}~\bibnamefont {Bao}},\ }\bibfield  {title} {\enquote {\bibinfo {title}
  {Manipulating polaritons at the extreme scale in van der {{Waals}}
  materials},}\ }\href {https://doi.org/10.1038/s42254-022-00472-0} {\bibfield
  {journal} {\bibinfo  {journal} {Nat Rev Phys}\ ,\ \bibinfo {pages} {1--17}}
  (\bibinfo {year} {2022})}\BibitemShut {NoStop}%
\bibitem [{\citenamefont {{Ordonez-Miranda}}\ \emph {et~al.}(2014)\citenamefont
  {{Ordonez-Miranda}}, \citenamefont {Tranchant}, \citenamefont {Kim},
  \citenamefont {Chalopin}, \citenamefont {Antoni},\ and\ \citenamefont
  {Volz}}]{ordonez-miranda_2014}%
  \BibitemOpen
  \bibfield  {author} {\bibinfo {author} {\bibfnamefont {J.}~\bibnamefont
  {{Ordonez-Miranda}}}, \bibinfo {author} {\bibfnamefont {L.}~\bibnamefont
  {Tranchant}}, \bibinfo {author} {\bibfnamefont {B.}~\bibnamefont {Kim}},
  \bibinfo {author} {\bibfnamefont {Y.}~\bibnamefont {Chalopin}}, \bibinfo
  {author} {\bibfnamefont {T.}~\bibnamefont {Antoni}},\ and\ \bibinfo {author}
  {\bibfnamefont {S.}~\bibnamefont {Volz}},\ }\bibfield  {title} {\enquote
  {\bibinfo {title} {Effects of anisotropy and size of polar nano thin films on
  their thermal conductivity due to surface phonon-polaritons},}\ }\href
  {https://doi.org/10.7567/apex.7.035201} {\bibfield  {journal} {\bibinfo
  {journal} {Appl. Phys. Express}\ }\textbf {\bibinfo {volume} {7}},\ \bibinfo
  {pages} {035201} (\bibinfo {year} {2014})}\BibitemShut {NoStop}%
\bibitem [{\citenamefont {{Ordonez-Miranda}}\ \emph {et~al.}(2013)\citenamefont
  {{Ordonez-Miranda}}, \citenamefont {Tranchant}, \citenamefont {Tokunaga},
  \citenamefont {Kim}, \citenamefont {Palpant}, \citenamefont {Chalopin},
  \citenamefont {Antoni},\ and\ \citenamefont {Volz}}]{ordonez-miranda_2013}%
  \BibitemOpen
  \bibfield  {author} {\bibinfo {author} {\bibfnamefont {J.}~\bibnamefont
  {{Ordonez-Miranda}}}, \bibinfo {author} {\bibfnamefont {L.}~\bibnamefont
  {Tranchant}}, \bibinfo {author} {\bibfnamefont {T.}~\bibnamefont {Tokunaga}},
  \bibinfo {author} {\bibfnamefont {B.}~\bibnamefont {Kim}}, \bibinfo {author}
  {\bibfnamefont {B.}~\bibnamefont {Palpant}}, \bibinfo {author} {\bibfnamefont
  {Y.}~\bibnamefont {Chalopin}}, \bibinfo {author} {\bibfnamefont
  {T.}~\bibnamefont {Antoni}},\ and\ \bibinfo {author} {\bibfnamefont
  {S.}~\bibnamefont {Volz}},\ }\bibfield  {title} {\enquote {\bibinfo {title}
  {Anomalous thermal conductivity by surface phonon-polaritons of polar nano
  thin films due to their asymmetric surrounding media},}\ }\href@noop {}
  {\bibfield  {journal} {\bibinfo  {journal} {J. Appl. Phys.}\ }\textbf
  {\bibinfo {volume} {113}},\ \bibinfo {pages} {084311} (\bibinfo {year}
  {2013})}\BibitemShut {NoStop}%
\bibitem [{\citenamefont {{Ordonez-Miranda}}\ \emph {et~al.}(2016)\citenamefont
  {{Ordonez-Miranda}}, \citenamefont {Tranchant}, \citenamefont {Joulain},
  \citenamefont {Ezzahri}, \citenamefont {Drevillon},\ and\ \citenamefont
  {Volz}}]{ordonez-miranda_2016}%
  \BibitemOpen
  \bibfield  {author} {\bibinfo {author} {\bibfnamefont {J.}~\bibnamefont
  {{Ordonez-Miranda}}}, \bibinfo {author} {\bibfnamefont {L.}~\bibnamefont
  {Tranchant}}, \bibinfo {author} {\bibfnamefont {K.}~\bibnamefont {Joulain}},
  \bibinfo {author} {\bibfnamefont {Y.}~\bibnamefont {Ezzahri}}, \bibinfo
  {author} {\bibfnamefont {J.}~\bibnamefont {Drevillon}},\ and\ \bibinfo
  {author} {\bibfnamefont {S.}~\bibnamefont {Volz}},\ }\bibfield  {title}
  {\enquote {\bibinfo {title} {Thermal energy transport in a surface
  phonon-polariton crystal},}\ }\href {https://doi.org/PhysRevB.93.035428}
  {\bibfield  {journal} {\bibinfo  {journal} {Phys. Rev. B}\ }\textbf {\bibinfo
  {volume} {93}},\ \bibinfo {pages} {035428} (\bibinfo {year}
  {2016})}\BibitemShut {NoStop}%
\bibitem [{\citenamefont {Tranchant}\ \emph {et~al.}(2019)\citenamefont
  {Tranchant}, \citenamefont {Hamamura}, \citenamefont {{Ordonez-Miranda}},
  \citenamefont {Yabuki}, \citenamefont {{Vega-Flick}}, \citenamefont
  {{Cervantes-Alvarez}}, \citenamefont {{Alvarado-Gil}}, \citenamefont {Volz},\
  and\ \citenamefont {Miyazaki}}]{tranchant_2019}%
  \BibitemOpen
  \bibfield  {author} {\bibinfo {author} {\bibfnamefont {L.}~\bibnamefont
  {Tranchant}}, \bibinfo {author} {\bibfnamefont {S.}~\bibnamefont {Hamamura}},
  \bibinfo {author} {\bibfnamefont {J.}~\bibnamefont {{Ordonez-Miranda}}},
  \bibinfo {author} {\bibfnamefont {T.}~\bibnamefont {Yabuki}}, \bibinfo
  {author} {\bibfnamefont {A.}~\bibnamefont {{Vega-Flick}}}, \bibinfo {author}
  {\bibfnamefont {F.}~\bibnamefont {{Cervantes-Alvarez}}}, \bibinfo {author}
  {\bibfnamefont {J.~J.}\ \bibnamefont {{Alvarado-Gil}}}, \bibinfo {author}
  {\bibfnamefont {S.}~\bibnamefont {Volz}},\ and\ \bibinfo {author}
  {\bibfnamefont {K.}~\bibnamefont {Miyazaki}},\ }\bibfield  {title} {\enquote
  {\bibinfo {title} {Two-{{Dimensional Phonon Polariton Heat Transport}}},}\
  }\href@noop {} {\bibfield  {journal} {\bibinfo  {journal} {Nano Lett.}\
  }\textbf {\bibinfo {volume} {19}},\ \bibinfo {pages} {6924--6930} (\bibinfo
  {year} {2019})}\BibitemShut {NoStop}%
\bibitem [{\citenamefont {Baudin}, \citenamefont {Voisin},\ and\ \citenamefont
  {Pla{\c c}ais}(2020)}]{baudin_2020}%
  \BibitemOpen
  \bibfield  {author} {\bibinfo {author} {\bibfnamefont {E.}~\bibnamefont
  {Baudin}}, \bibinfo {author} {\bibfnamefont {C.}~\bibnamefont {Voisin}},\
  and\ \bibinfo {author} {\bibfnamefont {B.}~\bibnamefont {Pla{\c c}ais}},\
  }\bibfield  {title} {\enquote {\bibinfo {title} {Hyperbolic {{Phonon
  Polariton Electroluminescence}} as an {{Electronic Cooling Pathway}}},}\
  }\href {https://doi.org/10.1002/adfm.201904783} {\bibfield  {journal}
  {\bibinfo  {journal} {Adv. Funct. Mater.}\ }\textbf {\bibinfo {volume}
  {30}},\ \bibinfo {pages} {1904783} (\bibinfo {year} {2020})}\BibitemShut
  {NoStop}%
\bibitem [{\citenamefont {Wu}\ \emph {et~al.}(2017)\citenamefont {Wu},
  \citenamefont {Varshney}, \citenamefont {Lee}, \citenamefont {Pang},
  \citenamefont {Roy},\ and\ \citenamefont {Luo}}]{wu_2017a}%
  \BibitemOpen
  \bibfield  {author} {\bibinfo {author} {\bibfnamefont {X.}~\bibnamefont
  {Wu}}, \bibinfo {author} {\bibfnamefont {V.}~\bibnamefont {Varshney}},
  \bibinfo {author} {\bibfnamefont {J.}~\bibnamefont {Lee}}, \bibinfo {author}
  {\bibfnamefont {Y.}~\bibnamefont {Pang}}, \bibinfo {author} {\bibfnamefont
  {A.~K.}\ \bibnamefont {Roy}},\ and\ \bibinfo {author} {\bibfnamefont
  {T.}~\bibnamefont {Luo}},\ }\bibfield  {title} {\enquote {\bibinfo {title}
  {How to characterize thermal transport capability of {{2D}} materials fairly?
  \textendash{} {{Sheet}} thermal conductance and the choice of thickness},}\
  }\href {https://doi.org/10.1016/j.cplett.2016.12.054} {\bibfield  {journal}
  {\bibinfo  {journal} {Chemical Physics Letters}\ }\textbf {\bibinfo {volume}
  {669}},\ \bibinfo {pages} {233--237} (\bibinfo {year} {2017})}\BibitemShut
  {NoStop}%
\bibitem [{\citenamefont {Togo}, \citenamefont {Chaput},\ and\ \citenamefont
  {Tanaka}(2015)}]{togo_2015}%
  \BibitemOpen
  \bibfield  {author} {\bibinfo {author} {\bibfnamefont {A.}~\bibnamefont
  {Togo}}, \bibinfo {author} {\bibfnamefont {L.}~\bibnamefont {Chaput}},\ and\
  \bibinfo {author} {\bibfnamefont {I.}~\bibnamefont {Tanaka}},\ }\bibfield
  {title} {\enquote {\bibinfo {title} {Distributions of phonon lifetimes in
  {{Brillouin}} zones},}\ }\href {https://doi.org/10.1103/PhysRevB.91.094306}
  {\bibfield  {journal} {\bibinfo  {journal} {Phys. Rev. B}\ }\textbf {\bibinfo
  {volume} {91}},\ \bibinfo {pages} {094306} (\bibinfo {year}
  {2015})}\BibitemShut {NoStop}%
\bibitem [{\citenamefont {Jarzembski}\ \emph {et~al.}(2020)\citenamefont
  {Jarzembski}, \citenamefont {Goldflam}, \citenamefont {Siddiqui},
  \citenamefont {Ruiz},\ and\ \citenamefont {Beechem}}]{jarzembski_2020}%
  \BibitemOpen
  \bibfield  {author} {\bibinfo {author} {\bibfnamefont {A.}~\bibnamefont
  {Jarzembski}}, \bibinfo {author} {\bibfnamefont {M.}~\bibnamefont
  {Goldflam}}, \bibinfo {author} {\bibfnamefont {A.}~\bibnamefont {Siddiqui}},
  \bibinfo {author} {\bibfnamefont {I.}~\bibnamefont {Ruiz}},\ and\ \bibinfo
  {author} {\bibfnamefont {T.~E.}\ \bibnamefont {Beechem}},\ }\bibfield
  {title} {\enquote {\bibinfo {title} {Enhancing {{Graphene Plasmonic Device
  Performance}} via its {{Dielectric Environment}}},}\ }\href
  {https://doi.org/10.1103/PhysRevApplied.14.034044} {\bibfield  {journal}
  {\bibinfo  {journal} {Phys. Rev. Appl.}\ }\textbf {\bibinfo {volume} {14}},\
  \bibinfo {pages} {034044} (\bibinfo {year} {09/16/ 2020})}\BibitemShut
  {NoStop}%
\bibitem [{\citenamefont {Novotny}\ and\ \citenamefont
  {Hecht}(2012)}]{novotny_2012}%
  \BibitemOpen
  \bibfield  {author} {\bibinfo {author} {\bibfnamefont {L.}~\bibnamefont
  {Novotny}}\ and\ \bibinfo {author} {\bibfnamefont {B.}~\bibnamefont
  {Hecht}},\ }\href@noop {} {\emph {\bibinfo {title} {Principles of
  Nano-Optics}}}\ (\bibinfo  {publisher} {{Cambridge university press}},\
  \bibinfo {year} {2012})\BibitemShut {NoStop}%
\bibitem [{\citenamefont {Borstel}\ and\ \citenamefont
  {Falge}(1977)}]{borstel_1977}%
  \BibitemOpen
  \bibfield  {author} {\bibinfo {author} {\bibfnamefont {G.}~\bibnamefont
  {Borstel}}\ and\ \bibinfo {author} {\bibfnamefont {H.~J.}\ \bibnamefont
  {Falge}},\ }\bibfield  {title} {\enquote {\bibinfo {title} {Surface
  phonon-polaritons at semi-infinite crystals},}\ }\href
  {https://doi.org/10.1002/pssb.2220830102} {\bibfield  {journal} {\bibinfo
  {journal} {Phys. Status Solidi B}\ }\textbf {\bibinfo {volume} {83}},\
  \bibinfo {pages} {11--45} (\bibinfo {year} {1977})}\BibitemShut {NoStop}%
\bibitem [{\citenamefont {Chen}(2007)}]{chen_2007a}%
  \BibitemOpen
  \bibfield  {author} {\bibinfo {author} {\bibfnamefont {D.-Z. A. D.-Z.~A.}\
  \bibnamefont {Chen}},\ }\emph {\bibinfo {title} {Energy Transmission through
  and along Thin Films Mediated by Surface Phonon-Polaritons}},\ \href@noop {}
  {\bibinfo {type} {Thesis}},\ \bibinfo  {school} {Massachusetts Institute of
  Technology} (\bibinfo {year} {2007})\BibitemShut {NoStop}%
\bibitem [{\citenamefont {Chen}, \citenamefont {Narayanaswamy},\ and\
  \citenamefont {Chen}(2005)}]{chen_2005a}%
  \BibitemOpen
  \bibfield  {author} {\bibinfo {author} {\bibfnamefont {D.-Z.~A.}\
  \bibnamefont {Chen}}, \bibinfo {author} {\bibfnamefont {A.}~\bibnamefont
  {Narayanaswamy}},\ and\ \bibinfo {author} {\bibfnamefont {G.}~\bibnamefont
  {Chen}},\ }\bibfield  {title} {\enquote {\bibinfo {title} {Surface
  phonon-polariton mediated thermal conductivity enhancement of amorphous thin
  films},}\ }\href {https://doi.org/10.1103/PhysRevB.72.155435} {\bibfield
  {journal} {\bibinfo  {journal} {Phys. Rev. B}\ }\textbf {\bibinfo {volume}
  {72}},\ \bibinfo {pages} {155435} (\bibinfo {year} {2005})}\BibitemShut
  {NoStop}%
\bibitem [{\citenamefont {Huber}\ \emph {et~al.}(2005)\citenamefont {Huber},
  \citenamefont {Ocelic}, \citenamefont {Kazantsev},\ and\ \citenamefont
  {Hillenbrand}}]{huber_2005}%
  \BibitemOpen
  \bibfield  {author} {\bibinfo {author} {\bibfnamefont {A.}~\bibnamefont
  {Huber}}, \bibinfo {author} {\bibfnamefont {N.}~\bibnamefont {Ocelic}},
  \bibinfo {author} {\bibfnamefont {D.}~\bibnamefont {Kazantsev}},\ and\
  \bibinfo {author} {\bibfnamefont {R.}~\bibnamefont {Hillenbrand}},\
  }\bibfield  {title} {\enquote {\bibinfo {title} {Near-field imaging of
  mid-infrared surface phonon polariton propagation},}\ }\href@noop {}
  {\bibfield  {journal} {\bibinfo  {journal} {Appl. Phys. Lett.}\ }\textbf
  {\bibinfo {volume} {87}},\ \bibinfo {pages} {081103} (\bibinfo {year}
  {2005})}\BibitemShut {NoStop}%
\bibitem [{\citenamefont {Lee}\ \emph {et~al.}(2014)\citenamefont {Lee},
  \citenamefont {Esfarjani}, \citenamefont {Luo}, \citenamefont {Zhou},
  \citenamefont {Tian},\ and\ \citenamefont {Chen}}]{lee_2014c}%
  \BibitemOpen
  \bibfield  {author} {\bibinfo {author} {\bibfnamefont {S.}~\bibnamefont
  {Lee}}, \bibinfo {author} {\bibfnamefont {K.}~\bibnamefont {Esfarjani}},
  \bibinfo {author} {\bibfnamefont {T.}~\bibnamefont {Luo}}, \bibinfo {author}
  {\bibfnamefont {J.}~\bibnamefont {Zhou}}, \bibinfo {author} {\bibfnamefont
  {Z.}~\bibnamefont {Tian}},\ and\ \bibinfo {author} {\bibfnamefont
  {G.}~\bibnamefont {Chen}},\ }\bibfield  {title} {\enquote {\bibinfo {title}
  {Resonant bonding leads to low lattice thermal conductivity},}\ }\href
  {https://doi.org/10.1038/ncomms4525} {\bibfield  {journal} {\bibinfo
  {journal} {Nat Commun}\ }\textbf {\bibinfo {volume} {5}},\ \bibinfo {pages}
  {3525} (\bibinfo {year} {2014})}\BibitemShut {NoStop}%
\bibitem [{\citenamefont {Luo}\ \emph {et~al.}(2013)\citenamefont {Luo},
  \citenamefont {Garg}, \citenamefont {Shiomi}, \citenamefont {Esfarjani},\
  and\ \citenamefont {Chen}}]{luo_2013a}%
  \BibitemOpen
  \bibfield  {author} {\bibinfo {author} {\bibfnamefont {T.}~\bibnamefont
  {Luo}}, \bibinfo {author} {\bibfnamefont {J.}~\bibnamefont {Garg}}, \bibinfo
  {author} {\bibfnamefont {J.}~\bibnamefont {Shiomi}}, \bibinfo {author}
  {\bibfnamefont {K.}~\bibnamefont {Esfarjani}},\ and\ \bibinfo {author}
  {\bibfnamefont {G.}~\bibnamefont {Chen}},\ }\bibfield  {title} {\enquote
  {\bibinfo {title} {Gallium arsenide thermal conductivity and optical phonon
  relaxation times from first-principles calculations},}\ }\href
  {https://doi.org/10.1209/0295-5075/101/16001} {\bibfield  {journal} {\bibinfo
   {journal} {EPL}\ }\textbf {\bibinfo {volume} {101}},\ \bibinfo {pages}
  {16001} (\bibinfo {year} {2013})}\BibitemShut {NoStop}%
\bibitem [{\citenamefont {Beechem}\ \emph {et~al.}(2016)\citenamefont
  {Beechem}, \citenamefont {McDonald}, \citenamefont {Fuller}, \citenamefont
  {Alec~Talin}, \citenamefont {Rost}, \citenamefont {Maria}, \citenamefont
  {Gaskins}, \citenamefont {Hopkins},\ and\ \citenamefont
  {Allerman}}]{beechem_2016a}%
  \BibitemOpen
  \bibfield  {author} {\bibinfo {author} {\bibfnamefont {T.~E.}\ \bibnamefont
  {Beechem}}, \bibinfo {author} {\bibfnamefont {A.~E.}\ \bibnamefont
  {McDonald}}, \bibinfo {author} {\bibfnamefont {E.~J.}\ \bibnamefont
  {Fuller}}, \bibinfo {author} {\bibfnamefont {A.}~\bibnamefont {Alec~Talin}},
  \bibinfo {author} {\bibfnamefont {C.~M.}\ \bibnamefont {Rost}}, \bibinfo
  {author} {\bibfnamefont {J.-P.}\ \bibnamefont {Maria}}, \bibinfo {author}
  {\bibfnamefont {J.~T.}\ \bibnamefont {Gaskins}}, \bibinfo {author}
  {\bibfnamefont {P.~E.}\ \bibnamefont {Hopkins}},\ and\ \bibinfo {author}
  {\bibfnamefont {A.~A.}\ \bibnamefont {Allerman}},\ }\bibfield  {title}
  {\enquote {\bibinfo {title} {Size dictated thermal conductivity of
  {{GaN}}},}\ }\href {https://doi.org/doi:http://dx.doi.org/10.1063/1.4962010}
  {\bibfield  {journal} {\bibinfo  {journal} {J. Appl. Phys.}\ }\textbf
  {\bibinfo {volume} {120}},\ \bibinfo {pages} {095104} (\bibinfo {year}
  {2016})}\BibitemShut {NoStop}%
\bibitem [{\citenamefont {Marconnet}, \citenamefont {Asheghi},\ and\
  \citenamefont {Goodson}(2013)}]{marconnet_2013}%
  \BibitemOpen
  \bibfield  {author} {\bibinfo {author} {\bibfnamefont {A.~M.}\ \bibnamefont
  {Marconnet}}, \bibinfo {author} {\bibfnamefont {M.}~\bibnamefont {Asheghi}},\
  and\ \bibinfo {author} {\bibfnamefont {K.~E.}\ \bibnamefont {Goodson}},\
  }\bibfield  {title} {\enquote {\bibinfo {title} {From the casimir limit to
  phononic crystals: 20 years of phonon transport studies using
  silicon-on-insulator technology},}\ }\href@noop {} {\bibfield  {journal}
  {\bibinfo  {journal} {J Heat Transf.}\ }\textbf {\bibinfo {volume} {135}},\
  \bibinfo {pages} {061601} (\bibinfo {year} {2013})}\BibitemShut {NoStop}%
\bibitem [{\citenamefont {Ziade}\ \emph {et~al.}(2017)\citenamefont {Ziade},
  \citenamefont {Yang}, \citenamefont {Brummer}, \citenamefont {Nothern},
  \citenamefont {Moustakas},\ and\ \citenamefont {Schmidt}}]{ziade_2017}%
  \BibitemOpen
  \bibfield  {author} {\bibinfo {author} {\bibfnamefont {E.}~\bibnamefont
  {Ziade}}, \bibinfo {author} {\bibfnamefont {J.}~\bibnamefont {Yang}},
  \bibinfo {author} {\bibfnamefont {G.}~\bibnamefont {Brummer}}, \bibinfo
  {author} {\bibfnamefont {D.}~\bibnamefont {Nothern}}, \bibinfo {author}
  {\bibfnamefont {T.}~\bibnamefont {Moustakas}},\ and\ \bibinfo {author}
  {\bibfnamefont {A.~J.}\ \bibnamefont {Schmidt}},\ }\bibfield  {title}
  {\enquote {\bibinfo {title} {Thickness dependent thermal conductivity of
  gallium nitride},}\ }\href@noop {} {\bibfield  {journal} {\bibinfo  {journal}
  {Appl. Phys. Lett.}\ }\textbf {\bibinfo {volume} {110}},\ \bibinfo {pages}
  {031903} (\bibinfo {year} {2017})}\BibitemShut {NoStop}%
\bibitem [{\citenamefont {Seyf}\ \emph {et~al.}(2017)\citenamefont {Seyf},
  \citenamefont {Yates}, \citenamefont {Bougher}, \citenamefont {Graham},
  \citenamefont {Cola}, \citenamefont {Detchprohm}, \citenamefont {Ji},
  \citenamefont {Kim}, \citenamefont {Dupuis}, \citenamefont {Lv},\ and\
  \citenamefont {Henry}}]{seyf_2017}%
  \BibitemOpen
  \bibfield  {author} {\bibinfo {author} {\bibfnamefont {H.~R.}\ \bibnamefont
  {Seyf}}, \bibinfo {author} {\bibfnamefont {L.}~\bibnamefont {Yates}},
  \bibinfo {author} {\bibfnamefont {T.~L.}\ \bibnamefont {Bougher}}, \bibinfo
  {author} {\bibfnamefont {S.}~\bibnamefont {Graham}}, \bibinfo {author}
  {\bibfnamefont {B.~A.}\ \bibnamefont {Cola}}, \bibinfo {author}
  {\bibfnamefont {T.}~\bibnamefont {Detchprohm}}, \bibinfo {author}
  {\bibfnamefont {M.-H.}\ \bibnamefont {Ji}}, \bibinfo {author} {\bibfnamefont
  {J.}~\bibnamefont {Kim}}, \bibinfo {author} {\bibfnamefont {R.}~\bibnamefont
  {Dupuis}}, \bibinfo {author} {\bibfnamefont {W.}~\bibnamefont {Lv}},\ and\
  \bibinfo {author} {\bibfnamefont {A.}~\bibnamefont {Henry}},\ }\bibfield
  {title} {\enquote {\bibinfo {title} {Rethinking phonons: {{The}} issue of
  disorder},}\ }\href {https://doi.org/10.1038/s41524-017-0052-9} {\bibfield
  {journal} {\bibinfo  {journal} {Npj Comput. Mater.}\ }\textbf {\bibinfo
  {volume} {3}},\ \bibinfo {pages} {49} (\bibinfo {year} {2017})}\BibitemShut
  {NoStop}%
\bibitem [{\citenamefont {Song}\ \emph {et~al.}(2021)\citenamefont {Song},
  \citenamefont {Perez}, \citenamefont {Esteves}, \citenamefont {Lundh},
  \citenamefont {Saltonstall}, \citenamefont {Beechem}, \citenamefont {Yang},
  \citenamefont {Ferri}, \citenamefont {Brown}, \citenamefont {Tang},
  \citenamefont {Maria}, \citenamefont {Snyder}, \citenamefont {Olsson},
  \citenamefont {Griffin}, \citenamefont {{Trolier-McKinstry}}, \citenamefont
  {Foley},\ and\ \citenamefont {Choi}}]{song_2021a}%
  \BibitemOpen
  \bibfield  {author} {\bibinfo {author} {\bibfnamefont {Y.}~\bibnamefont
  {Song}}, \bibinfo {author} {\bibfnamefont {C.}~\bibnamefont {Perez}},
  \bibinfo {author} {\bibfnamefont {G.}~\bibnamefont {Esteves}}, \bibinfo
  {author} {\bibfnamefont {J.~S.}\ \bibnamefont {Lundh}}, \bibinfo {author}
  {\bibfnamefont {C.~B.}\ \bibnamefont {Saltonstall}}, \bibinfo {author}
  {\bibfnamefont {T.~E.}\ \bibnamefont {Beechem}}, \bibinfo {author}
  {\bibfnamefont {J.~I.}\ \bibnamefont {Yang}}, \bibinfo {author}
  {\bibfnamefont {K.}~\bibnamefont {Ferri}}, \bibinfo {author} {\bibfnamefont
  {J.~E.}\ \bibnamefont {Brown}}, \bibinfo {author} {\bibfnamefont
  {Z.}~\bibnamefont {Tang}}, \bibinfo {author} {\bibfnamefont {J.-P.}\
  \bibnamefont {Maria}}, \bibinfo {author} {\bibfnamefont {D.~W.}\ \bibnamefont
  {Snyder}}, \bibinfo {author} {\bibfnamefont {R.~H.~I.}\ \bibnamefont
  {Olsson}}, \bibinfo {author} {\bibfnamefont {B.~A.}\ \bibnamefont {Griffin}},
  \bibinfo {author} {\bibfnamefont {S.~E.}\ \bibnamefont
  {{Trolier-McKinstry}}}, \bibinfo {author} {\bibfnamefont {B.~M.}\
  \bibnamefont {Foley}},\ and\ \bibinfo {author} {\bibfnamefont
  {S.}~\bibnamefont {Choi}},\ }\bibfield  {title} {\enquote {\bibinfo {title}
  {Thermal {{Conductivity}} of {{Aluminum Scandium Nitride}} for {{5G Mobile
  Applications}} and {{Beyond}}},}\ }\href
  {https://doi.org/10.1021/acsami.1c02912} {\bibfield  {journal} {\bibinfo
  {journal} {ACS Appl. Mater. Interfaces}\ }\textbf {\bibinfo {volume} {13}},\
  \bibinfo {pages} {19031--19041} (\bibinfo {year} {2021})}\BibitemShut
  {NoStop}%
\bibitem [{\citenamefont {Cheaito}\ \emph {et~al.}(2012)\citenamefont
  {Cheaito}, \citenamefont {Duda}, \citenamefont {Beechem}, \citenamefont
  {Hattar}, \citenamefont {Ihlefeld}, \citenamefont {Medlin}, \citenamefont
  {Rodriguez}, \citenamefont {Campion}, \citenamefont {Piekos},\ and\
  \citenamefont {Hopkins}}]{cheaito_2012}%
  \BibitemOpen
  \bibfield  {author} {\bibinfo {author} {\bibfnamefont {R.}~\bibnamefont
  {Cheaito}}, \bibinfo {author} {\bibfnamefont {J.~C.}\ \bibnamefont {Duda}},
  \bibinfo {author} {\bibfnamefont {T.~E.}\ \bibnamefont {Beechem}}, \bibinfo
  {author} {\bibfnamefont {K.}~\bibnamefont {Hattar}}, \bibinfo {author}
  {\bibfnamefont {J.~F.}\ \bibnamefont {Ihlefeld}}, \bibinfo {author}
  {\bibfnamefont {D.~L.}\ \bibnamefont {Medlin}}, \bibinfo {author}
  {\bibfnamefont {M.~A.}\ \bibnamefont {Rodriguez}}, \bibinfo {author}
  {\bibfnamefont {M.~J.}\ \bibnamefont {Campion}}, \bibinfo {author}
  {\bibfnamefont {E.~S.}\ \bibnamefont {Piekos}},\ and\ \bibinfo {author}
  {\bibfnamefont {P.~E.}\ \bibnamefont {Hopkins}},\ }\bibfield  {title}
  {\enquote {\bibinfo {title} {Experimental {{Investigation}} of {{Size
  Effects}} on the {{Thermal Conductivity}} of {{Silicon-Germanium Alloy Thin
  Films}}},}\ }\href@noop {} {\bibfield  {journal} {\bibinfo  {journal} {Phys.
  Rev. Lett.}\ }\textbf {\bibinfo {volume} {109}},\ \bibinfo {pages} {195901}
  (\bibinfo {year} {2012})}\BibitemShut {NoStop}%
\bibitem [{\citenamefont {Slutsky}\ and\ \citenamefont
  {Garland}(1959)}]{slutsky_1959}%
  \BibitemOpen
  \bibfield  {author} {\bibinfo {author} {\bibfnamefont {L.~J.}\ \bibnamefont
  {Slutsky}}\ and\ \bibinfo {author} {\bibfnamefont {C.~W.}\ \bibnamefont
  {Garland}},\ }\bibfield  {title} {\enquote {\bibinfo {title} {Elastic
  {{Constants}} of {{Indium Antimonide}} from 4.2 {{K}} to 300 {{K}}},}\ }\href
  {https://doi.org/10.1103/PhysRev.113.167} {\bibfield  {journal} {\bibinfo
  {journal} {Phys. Rev.}\ }\textbf {\bibinfo {volume} {113}},\ \bibinfo {pages}
  {167--169} (\bibinfo {year} {1959})}\BibitemShut {NoStop}%
\bibitem [{\citenamefont {Blakemore}(1982)}]{blakemore_1982}%
  \BibitemOpen
  \bibfield  {author} {\bibinfo {author} {\bibfnamefont {J.~S.}\ \bibnamefont
  {Blakemore}},\ }\bibfield  {title} {\enquote {\bibinfo {title}
  {Semiconducting and other major properties of gallium arsenide},}\ }\href
  {https://doi.org/10.1063/1.331665} {\bibfield  {journal} {\bibinfo  {journal}
  {Journal of Applied Physics}\ }\textbf {\bibinfo {volume} {53}},\ \bibinfo
  {pages} {R123--R181} (\bibinfo {year} {1982})}\BibitemShut {NoStop}%
\bibitem [{\citenamefont {Wright}(1997)}]{wright_1997}%
  \BibitemOpen
  \bibfield  {author} {\bibinfo {author} {\bibfnamefont {A.~F.}\ \bibnamefont
  {Wright}},\ }\bibfield  {title} {\enquote {\bibinfo {title} {Elastic
  properties of zinc-blende and wurtzite {{AlN}}, {{GaN}}, and {{InN}}},}\
  }\href@noop {} {\bibfield  {journal} {\bibinfo  {journal} {J. Appl. Phys.}\
  }\textbf {\bibinfo {volume} {82}},\ \bibinfo {pages} {2833} (\bibinfo {year}
  {1997})}\BibitemShut {NoStop}%
\bibitem [{\citenamefont {Beechem}\ and\ \citenamefont
  {Graham}(2008)}]{beechem_2008a}%
  \BibitemOpen
  \bibfield  {author} {\bibinfo {author} {\bibfnamefont {T.}~\bibnamefont
  {Beechem}}\ and\ \bibinfo {author} {\bibfnamefont {S.}~\bibnamefont
  {Graham}},\ }\bibfield  {title} {\enquote {\bibinfo {title} {Temperature and
  doping dependence of phonon lifetimes and decay pathways in {{GaN}}},}\
  }\href {https://doi.org/10.1063/1.2912819} {\bibfield  {journal} {\bibinfo
  {journal} {J. Appl. Phys.}\ }\textbf {\bibinfo {volume} {103}},\ \bibinfo
  {pages} {093507} (\bibinfo {year} {2008})}\BibitemShut {NoStop}%
\bibitem [{\citenamefont {Yang}\ \emph {et~al.}(2020)\citenamefont {Yang},
  \citenamefont {Feng}, \citenamefont {Kang}, \citenamefont {Hu}, \citenamefont
  {Li},\ and\ \citenamefont {Ruan}}]{yang_2020a}%
  \BibitemOpen
  \bibfield  {author} {\bibinfo {author} {\bibfnamefont {X.}~\bibnamefont
  {Yang}}, \bibinfo {author} {\bibfnamefont {T.}~\bibnamefont {Feng}}, \bibinfo
  {author} {\bibfnamefont {J.~S.}\ \bibnamefont {Kang}}, \bibinfo {author}
  {\bibfnamefont {Y.}~\bibnamefont {Hu}}, \bibinfo {author} {\bibfnamefont
  {J.}~\bibnamefont {Li}},\ and\ \bibinfo {author} {\bibfnamefont
  {X.}~\bibnamefont {Ruan}},\ }\bibfield  {title} {\enquote {\bibinfo {title}
  {Observation of strong higher-order lattice anharmonicity in {{Raman}} and
  infrared spectra},}\ }\href {https://doi.org/10.1103/PhysRevB.101.161202}
  {\bibfield  {journal} {\bibinfo  {journal} {Phys. Rev. B}\ }\textbf {\bibinfo
  {volume} {101}},\ \bibinfo {pages} {161202} (\bibinfo {year}
  {2020})}\BibitemShut {NoStop}%
\bibitem [{\citenamefont {Liarokapis}\ and\ \citenamefont
  {Anastassakis}(1984)}]{liarokapis_1984}%
  \BibitemOpen
  \bibfield  {author} {\bibinfo {author} {\bibfnamefont {E.}~\bibnamefont
  {Liarokapis}}\ and\ \bibinfo {author} {\bibfnamefont {E.}~\bibnamefont
  {Anastassakis}},\ }\bibfield  {title} {\enquote {\bibinfo {title} {Light
  scattering of {{InSb}} at high temperatures},}\ }\href
  {https://doi.org/10.1103/PhysRevB.30.2270} {\bibfield  {journal} {\bibinfo
  {journal} {Phys. Rev. B}\ }\textbf {\bibinfo {volume} {30}},\ \bibinfo
  {pages} {2270--2272} (\bibinfo {year} {1984})}\BibitemShut {NoStop}%
\bibitem [{\citenamefont {Verma}, \citenamefont {Abbi},\ and\ \citenamefont
  {Jain}(1995)}]{verma_1995}%
  \BibitemOpen
  \bibfield  {author} {\bibinfo {author} {\bibfnamefont {P.}~\bibnamefont
  {Verma}}, \bibinfo {author} {\bibfnamefont {S.~C.}\ \bibnamefont {Abbi}},\
  and\ \bibinfo {author} {\bibfnamefont {K.~P.}\ \bibnamefont {Jain}},\
  }\bibfield  {title} {\enquote {\bibinfo {title} {Raman-scattering probe of
  anharmonic effects in {{GaAs}}},}\ }\href
  {https://doi.org/10.1103/PhysRevB.51.16660} {\bibfield  {journal} {\bibinfo
  {journal} {Phys. Rev. B}\ }\textbf {\bibinfo {volume} {51}},\ \bibinfo
  {pages} {16660--16667} (\bibinfo {year} {1995})}\BibitemShut {NoStop}%
\bibitem [{\citenamefont {Carlson}, \citenamefont {Slack},\ and\ \citenamefont
  {Silverman}(1965)}]{carlson_1965}%
  \BibitemOpen
  \bibfield  {author} {\bibinfo {author} {\bibfnamefont {R.~O.}\ \bibnamefont
  {Carlson}}, \bibinfo {author} {\bibfnamefont {G.~A.}\ \bibnamefont {Slack}},\
  and\ \bibinfo {author} {\bibfnamefont {S.~J.}\ \bibnamefont {Silverman}},\
  }\bibfield  {title} {\enquote {\bibinfo {title} {Thermal {{Conductivity}} of
  {{GaAs}} and {{GaAs1}}- x {{P}} x {{Laser Semiconductors}}},}\ }\href@noop {}
  {\bibfield  {journal} {\bibinfo  {journal} {J. Appl. Phys.}\ }\textbf
  {\bibinfo {volume} {36}},\ \bibinfo {pages} {505--507} (\bibinfo {year}
  {1965})}\BibitemShut {NoStop}%
\bibitem [{\citenamefont {Wu}\ \emph {et~al.}(2020)\citenamefont {Wu},
  \citenamefont {{Ordonez-Miranda}}, \citenamefont {Gluchko}, \citenamefont
  {Anufriev}, \citenamefont {Meneses}, \citenamefont {Del~Campo}, \citenamefont
  {Volz},\ and\ \citenamefont {Nomura}}]{wu_2020}%
  \BibitemOpen
  \bibfield  {author} {\bibinfo {author} {\bibfnamefont {Y.}~\bibnamefont
  {Wu}}, \bibinfo {author} {\bibfnamefont {J.}~\bibnamefont
  {{Ordonez-Miranda}}}, \bibinfo {author} {\bibfnamefont {S.}~\bibnamefont
  {Gluchko}}, \bibinfo {author} {\bibfnamefont {R.}~\bibnamefont {Anufriev}},
  \bibinfo {author} {\bibfnamefont {D.~D.~S.}\ \bibnamefont {Meneses}},
  \bibinfo {author} {\bibfnamefont {L.}~\bibnamefont {Del~Campo}}, \bibinfo
  {author} {\bibfnamefont {S.}~\bibnamefont {Volz}},\ and\ \bibinfo {author}
  {\bibfnamefont {M.}~\bibnamefont {Nomura}},\ }\bibfield  {title} {\enquote
  {\bibinfo {title} {Enhanced thermal conduction by surface
  phonon-polaritons},}\ }\href {https://doi.org/10.1126/sciadv.abb4461}
  {\bibfield  {journal} {\bibinfo  {journal} {Sci. Adv.}\ }\textbf {\bibinfo
  {volume} {6}},\ \bibinfo {pages} {eabb4461} (\bibinfo {year}
  {2020})}\BibitemShut {NoStop}%
\bibitem [{\citenamefont {Chen}\ and\ \citenamefont {Chen}(2010)}]{chen_2010b}%
  \BibitemOpen
  \bibfield  {author} {\bibinfo {author} {\bibfnamefont {D.-Z.}\ \bibnamefont
  {Chen}}\ and\ \bibinfo {author} {\bibfnamefont {G.}~\bibnamefont {Chen}},\
  }\bibfield  {title} {\enquote {\bibinfo {title} {Heat flow in thin films via
  surface phonon-polaritons},}\ }\href {https://doi.org/10.5098/hmt.v1.2.3005}
  {\bibfield  {journal} {\bibinfo  {journal} {Front. Heat Mass Transf. FHMT}\
  }\textbf {\bibinfo {volume} {1}},\ \bibinfo {pages} {023005} (\bibinfo {year}
  {2010})}\BibitemShut {NoStop}%
\end{thebibliography}
\end{document}